\documentclass{article}
\usepackage{spconf,amsmath,amssymb,graphicx,booktabs,hyperref}
\hypersetup{hidelinks}
\usepackage{microtype}
\title{VoiceWeaver: Staged Learning of Structured Controls\\
for Expressive Speech and Sound-Event Generation}
\name{Xiaosu Su, Yun Cao, Yiping Ni, and Xiaowei Yi}
\address{Institute of Information Engineering, Chinese Academy of Sciences, Beijing, China\\School of Cyber Security, University of Chinese Academy of Sciences, Beijing, China}
\begin{document}
\maketitle
\begin{abstract}
Adding expressive and environmental controls to a speech generator requires learning heterogeneous attributes without losing earlier capabilities. VoiceWeaver addresses this problem with structured label prefixes and staged emotion--tone--event training in a shared text--audio model. Replay-based distillation retains earlier predictions, while embedding decorrelation and attribute dropout regularize conditioning. Evaluation separates single-attribute correctness, joint emotion--event generation, and three-attribute correctness. Rounded to whole percentage points, emotion accuracy is 86\% in Chinese and 83\% in English. Chinese emotion--event joint accuracy is about 70\%, versus 56\% for mixed-task training and 48\% for Ming-omni-tts; English joint accuracy is 66\%. Three-attribute accuracy is 59\% in Chinese and 55\% in English on 500 samples each (57\% pooled). Text error rates exceed those of the external TTS baselines. Audio samples are available at \url{https://anonymous.4open.science/api/repo/VoiceWeaver1-4D88/file/index.html}.
\end{abstract}

\section{Introduction}
Speech for narration, games, and audiovisual production must coordinate text, delivery, and acoustic scene. Emotion is a global affective state, tone is more text-dependent, and events add non-speech acoustics. Their distinct supervision and acoustic boundaries make it difficult to add controls while retaining linguistic content and earlier capabilities. Accepting multiple conditions alone does not establish joint realization.

Codec language models provide a shared discrete-token formulation for speech generation~\cite{valle,valler,seedtts,cosyvoice,cosyvoice2,fireredtts,sparktts,megatts2,moshi,survey}. PromptTTS and related systems condition style using text descriptions~\cite{prompttts,prompttts2,promptstyle,prompttag,styletts,indextts2,instructtts}, while factorized methods separate speaker, content, and style representations~\cite{naturalspeech2,naturalspeech3,speechtokenizer,controlspeech,speechtriplenet,hsu2019}. VoiceLDM and Speak in the Scene model environmental context~\cite{voiceldm,speakscene}; Ming-omni-tts supports speech and event prompts~\cite{ming}. Unlike reference-audio or open-ended prompt control, our label interface supplies a small, explicit vocabulary. The central question is whether these heterogeneous labels can be acquired sequentially without losing earlier controls, and whether requested emotion and event are realized together.

VoiceWeaver couples a structured control interface with an emotion--tone--event curriculum. Label prefixes separate requested attributes from linguistic content within one token sequence. The curriculum first learns stable affective cues, then adds delivery styles and environmental events. Distillation on replayed examples targets forgetting across stages; embedding decorrelation targets overlap between control groups; attribute dropout trains the model to accept incomplete label sets. A low-frame-rate tokenizer represents speech and events without changing this interface.

The contribution is a training and evaluation framework for this setting, rather than a new prefix mechanism or distillation loss. We test staged acquisition against mixed training and alternative orders, and separate bilingual attribute accuracy, emotion--event joint correctness, content fidelity, and perceptual coexistence. A pooled bilingual three-attribute test measures simultaneous correctness directly, complementing the larger emotion--event benchmark.

\section{VoiceWeaver}
\subsection{Structured conditioning and audio generation}
Given text $x$, a speaker token $s$, and optional attributes $z=(z_e,z_t,z_v)$, the model generates audio codes $a$ and reconstructs a waveform $y=D(a)$. Subscripts $e,t,v$ denote emotion, tone, and event. The conditioning sequence is
\begin{equation}
\begin{aligned}
c&=[p(z),s,\langle\mathrm{text}\rangle,x,\langle/\mathrm{text}\rangle],\\
p(z)&=[e_e(z_e),e_t(z_t),e_v(z_v)],
\end{aligned}
\label{eq:prefix}
\end{equation}
where $e_k$ denotes a learned label embedding; absent attributes are omitted. Labels specify global requests for an utterance, rather than time-aligned event locations. In this work, \emph{tone} refers to delivery styles such as gentle or hesitant, not lexical tone. The vocabulary contains six emotions, five tones, and twelve sound events.

For example, happy and rain request happy speech in a rainy scene; gentle adds a delivery style.

\begin{figure}[t]
\centering
\includegraphics[width=\columnwidth,trim=0bp 0bp 0bp 2bp,clip]{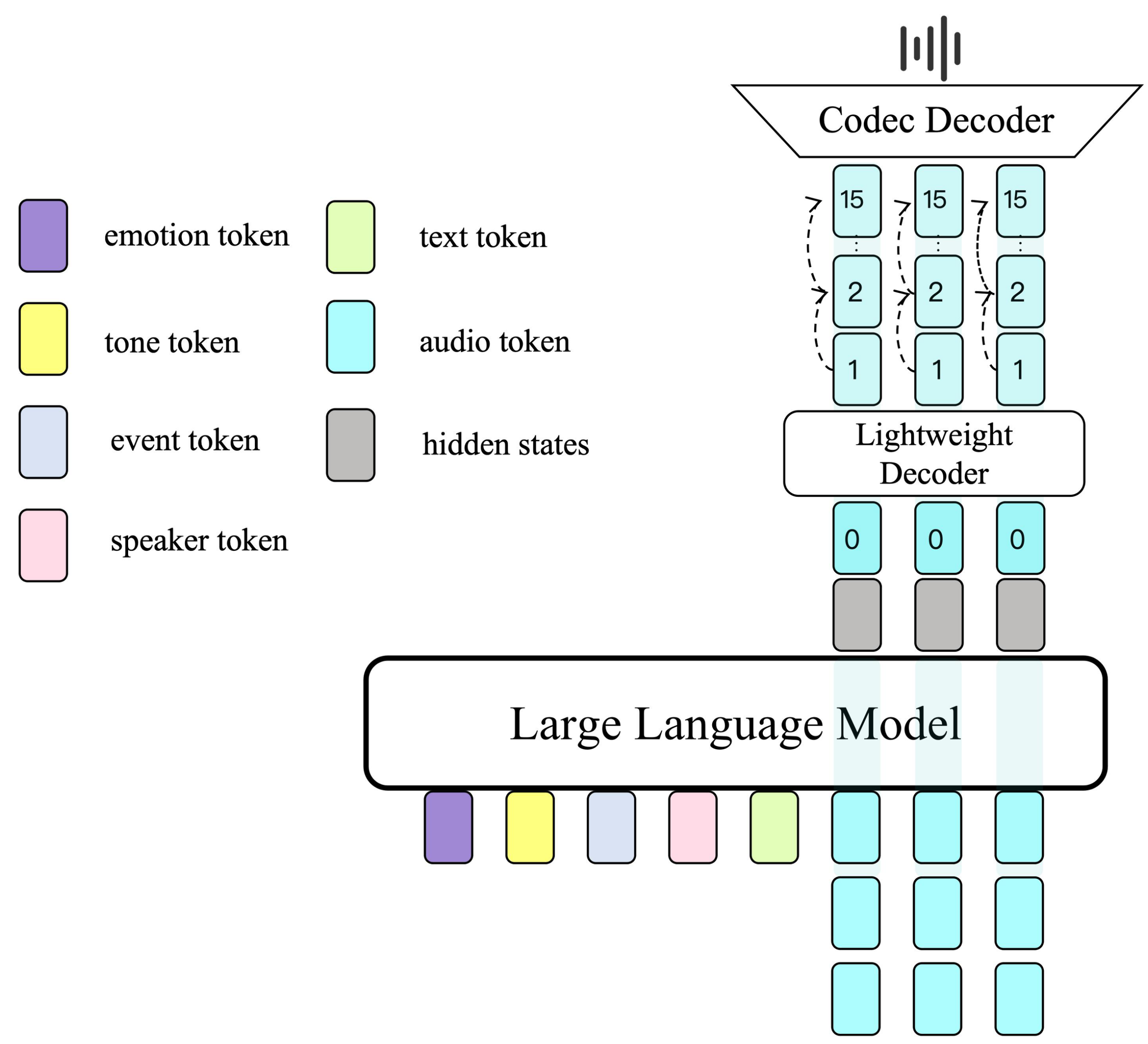}
\caption{VoiceWeaver architecture and token types. The autoregressive backbone predicts codebook 0; the lightweight decoder predicts codebooks 1--15 before codec reconstruction.}
\label{fig:architecture}
\end{figure}

VoiceWeaver-Tokenizer-12Hz encodes 16-kHz audio at 12.5 frames/s with 16 residual codebooks of 2,048 entries. Its encoder combines semantic and acoustic features before quantization, and its causal decoder reconstructs speech and events. A Qwen-scale 1.5B backbone predicts the semantic codebook; a 200M decoder predicts the remaining 15. Figures~\ref{fig:architecture} and~\ref{fig:stages} summarize the two paths.

\subsection{Staged learning with retention}
\begin{figure*}[t]
\centering
\includegraphics[width=0.72\textwidth,trim=45bp 390bp 10bp 400bp,clip]{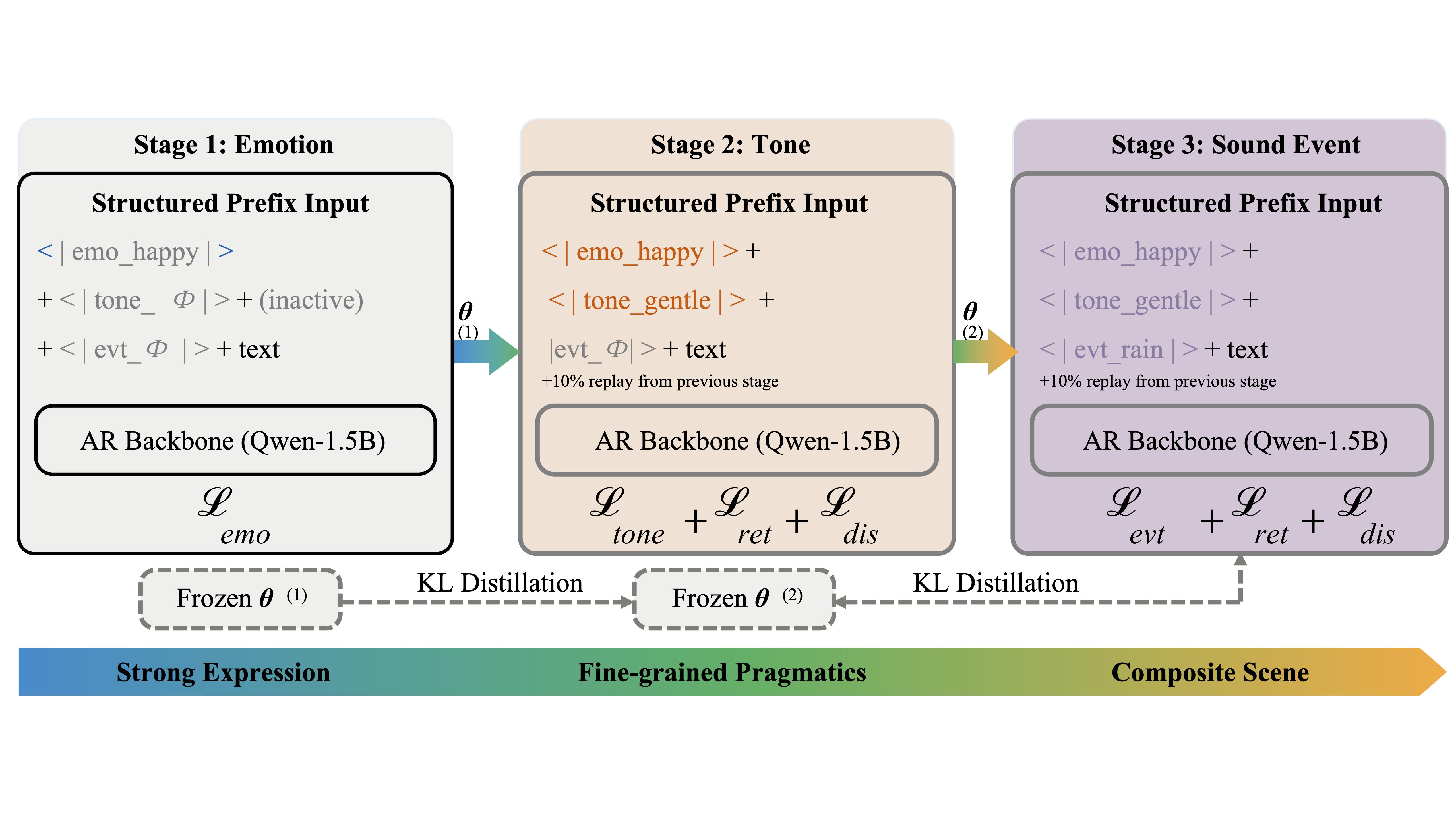}
\caption{Three-stage training. Emotion, tone, and sound-event controls are introduced in sequence; replay and frozen teachers retain earlier capabilities.}
\label{fig:stages}
\end{figure*}

The curriculum starts with affect, adds delivery style, then non-speech events; Figure~\ref{fig:stages} shows retention.

Let $L_{c0}$ and $L_{\mathrm{dec}}$ be cross-entropy losses for the first and remaining codebooks, and $L_{\mathrm{txt}}$ an auxiliary text loss. The base objective is
\begin{equation}
\mathcal L_{\mathrm{AR}}=2\bigl[(1-\alpha)L_{c0}+\alpha L_{\mathrm{dec}}\bigr]
+\lambda L_{\mathrm{txt}},
\label{eq:ar}
\end{equation}
with $\alpha=0.6$ and $\lambda=0.01$. At stage $r>1$, a frozen previous-stage model $\theta^{r-1}$ acts as a teacher on replay buffer $\mathcal R^{r-1}$. Retention penalizes changes in its predictive code distributions:
\begin{equation}
\mathcal L_{\mathrm{ret}}^r=
\sum_{u\in\mathcal R^{r-1}}\sum_{j,q}
D_{\mathrm{KL}}\!\left(P_{\theta^{r-1}}^{j,q}(u)\,\|\,P_{\theta^r}^{j,q}(u)\right),
\label{eq:ret}
\end{equation}
where $u$ is a replayed training example, $j$ indexes audio frames, $q$ indexes codebooks, and both models use the same teacher-forced context.

Here $q\in\{0,\ldots,15\}$, and each $P^{j,q}$ is a distribution over 2,048 code entries. Stage 2 uses the frozen emotion-stage model and emotion replay; stage 3 uses the frozen emotion--tone model and replay from the earlier stages. The distillation teacher is thus the preceding VoiceWeaver checkpoint, distinct from the IndexTTS2 teacher used to synthesize training data.

We apply a soft decorrelation penalty to label embeddings. Let $E_k\in\mathbb R^{d\times n_k}$ contain the $n_k$ labels of attribute $k$ as columns, and let $\bar E_k$ be centered across labels and column-wise $\ell_2$ normalized. We use $\mathcal L_{\mathrm{dis}}=\|\bar E_e^{\mathsf T}\bar E_t\|_F^2+\|\bar E_e^{\mathsf T}\bar E_v\|_F^2+\|\bar E_t^{\mathsf T}\bar E_v\|_F^2$, a soft constraint on conditioning embeddings rather than full audio-factor independence.
For each available attribute $k\in\{e,t,v\}$, dropout samples $m_k\sim\mathrm{Bernoulli}(0.8)$ independently; the label is omitted when $m_k=0$. Text and speaker conditions remain present. The stage objectives are
\begin{equation}
\mathcal L^1=\mathcal L_{\mathrm{AR}}^1,\qquad
\mathcal L^r=\mathcal L_{\mathrm{AR}}^r+\beta\mathcal L_{\mathrm{ret}}^r
+\gamma\mathcal L_{\mathrm{dis}},\quad r=2,3,
\label{eq:total}
\end{equation}
with $\beta=0.5$ and $\gamma=0.1$. Replay occupies 10\% of training samples in the later stages. We train for 80K, 50K, and 60K steps, respectively, using AdamW, learning rate $10^{-5}$, batch size 16, and 5,000 warmup steps.

\subsection{Training data}
Emotion and tone stages use 1,204,800 bilingual synthetic examples (783,120 Chinese and 421,680 English; approximately 3,500 h). An emotion-enhanced IndexTTS2 teacher supplies emotion data. Tone data combines LLM-generated, tone-matched text with teacher synthesis. Screening eleven candidate tones retains five with relatively recognizable acoustic cues: gentle, hesitant, coquettish, complaining, and resigned. Text dependence is reduced by this screening, but is not eliminated.

Tone screening uses twenty neutral sentences per category and three perceptibility judges.

The event stage uses 862,400 examples (552,000 Chinese and 310,400 English; approximately 2,500 h) from twelve ESC-50~\cite{esc50} categories. The 480 clips derive from 96 original recordings. Mixtures use a relative speech--event level range of 5--20 dB; the tokenizer uses approximately 10,000 h.

Event screening reconstructs candidate clips through the tokenizer and retains PANNs-recognizable categories. Short events are placed randomly; longer ones are cropped and aligned. Event data includes clean-event and emotion--event examples. PANNs also evaluates generated events, so that score is not independent of category selection.

\section{Experimental Protocol}
\subsection{Test sets and baselines}
Emotion evaluation uses 600 Chinese and 600 English test prompts, balanced over six emotions. Chinese tone evaluation uses 300 prompts, with 60 for each retained tone. The joint speech--scene test has 720 text--label combinations per language: six emotions crossed with twelve events and ten texts per combination. A separate three-attribute evaluation contains 500 Chinese and 500 English samples. Test texts in the original benchmarks do not overlap the training corpus. Event clips are drawn from the 96-recording source pool; recording-level separation from training was not documented, so the original joint benchmark does not measure generalization to unseen event recordings. We round the original classification percentages to whole points because available aggregates do not specify the generation count or averaging protocol behind each rate.

We compare official IndexTTS2 (reference-audio control), CosyVoice~2 (natural-language instructions), and Ming-omni-tts (one-to-one event prompts). All systems receive the same texts; external systems use default recipes and VoiceWeaver uses fixed decoding. Scale, data, and interface differ, so these are system-level comparisons.

E--V--T and T--E--V change stage order; mixed-task training removes staging. Component ablations remove KL retention, embedding decorrelation, or label dropout. The full model uses 190K steps and 10\% replay in later stages; per-attribute exposure in alternatives is not established by these totals.

\subsection{Automatic and perceptual evaluation}
Emotion accuracy uses emotion2vec~\cite{emotion2vec} (92.3\% ground-truth upper bound); event accuracy uses PANNs~\cite{panns} (89.7\% upper bound). Tone accuracy uses majority votes from three annotators, with Qwen3-Omni as a cross-check. FunASR Paraformer computes transcript error. We report CER for Chinese, WER for English, and naturalness MOS alongside attribute correctness. Let $b_i$ indicate correct emotion and event, and $t_i$ correct tone, on sample $i$. We define
\begin{equation}
\begin{aligned}
\mathrm{JointAcc}&=\frac{1}{N}\sum_i b_i,\\
\mathrm{MC}&=\frac{1}{N}\sum_i b_i\,\mathbf1[\mathrm{CER}_i<0.08],
\end{aligned}
\label{eq:joint}
\end{equation}
where multimodal consistency (MC) adds a content constraint. Three-attribute accuracy is $N_3^{-1}\sum_{i=1}^{N_3}b_it_i$, with $N_3=500$ for each language and 1,000 pooled; it uses the same emotion2vec, PANNs, and three-annotator majority decisions described above. MC uses an 8\% character-error threshold for both languages; standalone English intelligibility is WER. Automatic event scores inherit the limitations of PANNs, which also screened event categories.

For MOS and coexistence naturalness (CN), each system contributes 60 samples per language, each rated by at least five listeners on a 1--5 scale. There are 15 Chinese and 10 English listeners; order is randomized and identity hidden. Reported Krippendorff's $\alpha$ is 0.70 for MOS and 0.62 for CN. Each listener rates 30 matched pairs. Preferences are descriptive because listener- and clip-level records are unavailable for repeated-measures analysis.

\section{Results and Analysis}
\subsection{Expressive control and text fidelity}
VoiceWeaver reaches about 86\%/83\% Chinese/English emotion accuracy and 70\% Chinese tone accuracy (Table~\ref{tab:expressive}). Its CER/WER of 4.31\%/4.68\% trails the external TTS systems; the no-label model is already at 4.18\%/4.53\%, so this gap cannot be attributed solely to conditioning. Training data also differ greatly: about 6,000 h for VoiceWeaver versus 55K h for IndexTTS2 and 166K h for CosyVoice~2.

\begin{table}[t]
\centering
\caption{Expressive control and intelligibility (\%). Classification accuracies are rounded to whole points; CER/WER retain two decimals. Tone is evaluated in Chinese; ``--'' denotes no matched tone result.}
\label{tab:expressive}
\small\setlength{\tabcolsep}{3.5pt}
\begin{tabular}{@{}lrrrrr@{}}
\toprule
&\multicolumn{2}{c}{Emotion $\uparrow$}&Tone $\uparrow$&CER $\downarrow$&WER $\downarrow$\\
Model&ZH&EN&ZH&ZH&EN\\
\midrule
No labels&19&17&23&4.18&4.53\\
IndexTTS2&83&80&--&1.88&\textbf{2.34}\\
CosyVoice 2&75&72&--&\textbf{1.52}&2.57\\
Ming-omni-tts&77&75&--&2.14&2.82\\
VoiceWeaver&\textbf{86}&\textbf{83}&\textbf{70}&4.31&4.68\\
\bottomrule
\end{tabular}
\end{table}

Replacing tone-matched texts with random texts reduces tone accuracy from about 70\% to 64\%, indicating substantial text dependence. Chinese emotion MOS is 4.24 for VoiceWeaver versus 4.16/4.22 for IndexTTS2/CosyVoice~2; these means do not establish significance.

\subsection{Joint speech--scene generation}
VoiceWeaver improves reported emotion--event joint accuracy over Ming-omni-tts from 48\% to 70\% in Chinese and 46\% to 66\% in English (Table~\ref{tab:joint}). CN rises from 3.74 to 4.21 and from 3.68 to 4.14, respectively. These system-level differences combine training, interface, and architecture effects.

\begin{table}[t]
\centering
\caption{Emotion--event joint generation. Event, Joint, and MC are rounded percentages; CN is rated from 1 to 5. All columns: higher is better. Joint and MC do not include tone correctness.}
\label{tab:joint}
\small\setlength{\tabcolsep}{4.5pt}
\begin{tabular}{@{}llrrrr@{}}
\toprule
Lang.&Model&Event&Joint&MC&CN\\
\midrule
ZH&Ming-omni-tts&63&48&46&3.74\\
&VoiceWeaver&\textbf{79}&\textbf{70}&\textbf{67}&\textbf{4.21}\\
\midrule
EN&Ming-omni-tts&62&46&43&3.68\\
&VoiceWeaver&\textbf{76}&\textbf{66}&\textbf{64}&\textbf{4.14}\\
\bottomrule
\end{tabular}
\end{table}

Paired listening favors VoiceWeaver for emotion, event clarity, and scene naturalness (about 68\%, 72\%, and 75\%). These descriptive rates complement classifier scores but do not resolve listener dependence or interface confounds.

\subsection{Training order and component ablations}
The selected order reaches about 70\% joint accuracy, versus 63\% for E--V--T, 59\% for T--E--V, and 56\% for mixed-task training (Table~\ref{tab:ablation}). The comparison favors this curriculum within the reported setup, but per-attribute exposure is not audited.

\begin{table}[t]
\centering
\caption{Training orders and ablations (rounded \%). E/T/V: emotion/tone/event. Joint is Chinese emotion--event accuracy; single-attribute emotion and tone use their respective test sets.}
\label{tab:ablation}
\small\setlength{\tabcolsep}{4pt}
\begin{tabular}{@{}lrrrr@{}}
\toprule
Training / variant&Emo.&Tone&Event&Joint\\
\midrule
Full (E--T--V)&\textbf{86}&\textbf{70}&\textbf{79}&\textbf{70}\\
E--V--T&84&67&75&63\\
T--E--V&84&65&74&59\\
Mixed tasks&82&62&72&56\\
\midrule
w/o retention distillation&83&67&76&63\\
w/o decorrelation&83&68&75&64\\
w/o attribute dropout&84&69&76&66\\
Standard-teacher data&82&69&77&64\\
\bottomrule
\end{tabular}
\end{table}

Removing retention, decorrelation, or dropout lowers joint accuracy by roughly 6, 5, or 4 pp. Standard rather than enhanced emotion-teacher data lowers it to 64\%; this changes supervision, not just a loss term. These comparisons do not quantify run-to-run variance.

Regularization is non-monotonic: reported joint accuracy is about 64\% at $\gamma=0$, 70\% at 0.1, and 66\% at 0.3.

\subsection{Three-attribute control and stability}
Among 500 samples per language requested with emotion, tone, and event labels, 59\% of Chinese and 55\% of English outputs satisfy all three simultaneously (57\% pooled). On these same outputs, emotion, tone, and event accuracies are 82\%, 71\%, and 77\% in Chinese, and 80\%, 69\%, and 74\% in English, respectively. This directly tests three-way correctness, unlike the emotion--event benchmark or anchor-condition diagnostics. Across 100 fixed texts, the reported SD of a fixed attribute's accuracy stays near 1 pp when another attribute is varied under one anchor setting. No matched three-attribute baseline on this set is available.

\subsection{Scope and limitations}
Evidence covers bilingual closed-set three-attribute control and emotion--event composition, but tone is text-sensitive and weak or continuous events can be masked. The three-way test lacks a matched baseline. Event recording holdout remains unverified, and PANNs was used for both category screening and evaluation. Synthetic data, a small event library, and the text-fidelity gap limit generalization.

\section{Conclusion}
VoiceWeaver learns emotion, tone, and event controls sequentially in a shared text--audio generator. Structured prefixes, replay distillation, and conditioning regularizers improve emotion--event composition in the reported ablations. On the closed-set benchmarks, emotion--event joint accuracy reaches 70\%/66\% in Chinese/English; three-attribute correctness reaches 59\%/55\% on 500 prompts per language. These findings support staged training within this setup, although text fidelity trails external TTS baselines. PANNs screens and evaluates event categories, and recording-level holdout remains unverified. Independent event scoring, recording-disjoint tests, and matched three-attribute baselines are needed before claiming broader generalization.


\clearpage
\begin{thebibliography}{29}
\bibitem{valle} C. Wang et al., ``Neural codec language models are zero-shot text to speech synthesizers,'' arXiv:2301.02111, 2023.
\bibitem{valler} Z. Zhang et al., ``Speak foreign languages with your own voice: Cross-lingual neural codec language modeling,'' arXiv:2303.03926, 2023.
\bibitem{seedtts} P. Anastassiou et al., ``Seed-TTS: A family of high-quality versatile speech generation models,'' arXiv:2406.02430, 2024.
\bibitem{cosyvoice2} Z. Du et al., ``CosyVoice 2: Scalable streaming speech synthesis with large language models,'' arXiv:2412.10117, 2024.
\bibitem{fireredtts} H.-H. Guo et al., ``FireRedTTS: A foundation text-to-speech framework for industry-level generative speech applications,'' arXiv:2409.03283, 2024.
\bibitem{sparktts} X. Wang et al., ``Spark-TTS: An efficient LLM-based text-to-speech model with single-stream decoupled speech tokens,'' arXiv:2503.01710, 2025.
\bibitem{prompttts} Z. Guo, Y. Leng, Y. Wu, S. Zhao, and X. Tan, ``PromptTTS: Controllable text-to-speech with text descriptions,'' arXiv:2211.12171, 2023.
\bibitem{prompttts2} Y. Leng et al., ``PromptTTS 2: Describing and generating voices with text prompt,'' arXiv:2309.02285, 2023.
\bibitem{promptstyle} G. Liu et al., ``PromptStyle: Controllable style transfer for text-to-speech with natural language descriptions,'' arXiv:2305.19522, 2023.
\bibitem{indextts2} S. Zhou et al., ``IndexTTS2: A breakthrough in emotionally expressive and duration-controlled auto-regressive zero-shot text-to-speech,'' arXiv:2506.21619, 2025.
\bibitem{instructtts} D. Yang et al., ``InstructTTS: Modelling expressive TTS in discrete latent space with natural language style prompt,'' arXiv:2301.13662, 2023.
\bibitem{naturalspeech2} K. Shen et al., ``NaturalSpeech 2: Latent diffusion models are natural and zero-shot speech and singing synthesizers,'' arXiv:2304.09116, 2023.
\bibitem{naturalspeech3} Z. Ju et al., ``NaturalSpeech 3: Zero-shot speech synthesis with factorized codec and diffusion models,'' in \emph{Proc. ICML}, 2024.
\bibitem{speechtokenizer} X. Zhang et al., ``SpeechTokenizer: Unified speech tokenizer for speech large language models,'' arXiv:2308.16692, 2023.
\bibitem{controlspeech} S. Ji et al., ``ControlSpeech: Towards simultaneous and independent zero-shot speaker cloning and zero-shot language style control,'' arXiv:2406.01205, 2024.
\bibitem{speechtriplenet} H. Lu, X. Wu, Z. Wu, and H. Meng, ``SpeechTripleNet: End-to-end disentangled speech representation learning for content, timbre and prosody,'' in \emph{Proc. ACM Multimedia}, 2023, pp. 2829--2837.
\bibitem{hsu2019} W.-N. Hsu et al., ``Disentangling correlated speaker and noise for speech synthesis via data augmentation and adversarial factorization,'' in \emph{Proc. ICASSP}, 2019, pp. 5901--5905.
\bibitem{voiceldm} Y. Lee, I. Yeon, J. Nam, and J. S. Chung, ``VoiceLDM: Text-to-speech with environmental context,'' in \emph{Proc. ICASSP}, 2024, pp. 12566--12571.
\bibitem{speakscene} M. Kim, S.-W. Chung, Y. Ji, H.-G. Kang, and M.-S. Choi, ``Speak in the scene: Diffusion-based acoustic scene transfer toward immersive speech generation,'' in \emph{Proc. Interspeech}, 2024, pp. 4883--4887.
\bibitem{ming} inclusionAI, ``Ming-omni-tts: Unified audio generation with precise control,'' 2025. [Online]. Available: \url{https://github.com/inclusionAI/Ming-omni-tts}.
\bibitem{esc50} K. J. Piczak, ``ESC: Dataset for environmental sound classification,'' in \emph{Proc. ACM Multimedia}, 2015, pp. 1015--1018.
\bibitem{panns} Q. Kong, Y. Cao, T. Iqbal, Y. Wang, W. Wang, and M. D. Plumbley, ``PANNs: Large-scale pretrained audio neural networks for audio pattern recognition,'' \emph{IEEE/ACM Trans. Audio, Speech, Lang. Process.}, vol. 28, pp. 2880--2894, 2020.
\bibitem{emotion2vec} Z. Ma et al., ``emotion2vec: Self-supervised pre-training for speech emotion representation,'' arXiv:2312.15185, 2023.
\bibitem{cosyvoice} Z. Du et al., ``CosyVoice: A scalable multilingual zero-shot text-to-speech synthesizer based on supervised semantic tokens,'' arXiv:2407.05407, 2024.
\bibitem{prompttag} M. Kim et al., ``Expressive text-to-speech using style tag,'' in \emph{Proc. Interspeech}, 2021, pp. 4663--4667.
\bibitem{styletts} Y. A. Li et al., ``StyleTTS: A style-based generative model for natural and diverse text-to-speech synthesis,'' \emph{IEEE J. Sel. Top. Signal Process.}, vol. 19, no. 1, pp. 283--296, 2025.
\bibitem{moshi} A. Défossez et al., ``Moshi: A speech-text foundation model for real-time dialogue,'' arXiv:2410.00037, 2024.
\bibitem{megatts2} Z. Jiang et al., ``Mega-TTS 2: Boosting prompting mechanisms for zero-shot speech synthesis,'' arXiv:2307.07218, 2023.
\bibitem{survey} T. Xie et al., ``Towards controllable speech synthesis in the era of large language models: A systematic survey,'' arXiv:2412.06602, 2024.
\end{thebibliography}
\end{document}